# Anomalous Thermodynamic Cost of Clock Synchronization

Cheng Yang[1], Jiteng Sheng[1,2,*], and Haibin Wu[1,2,3,*]

[1]State Key Laboratory of Precision Spectroscopy, Institute of Quantum Science and Precision Measurement, East China Normal University, Shanghai 200062, China

[2]Collaborative Innovation Center of Extreme Optics, Shanxi University, Taiyuan 030006, China

[3]Shanghai Research Center for Quantum Sciences, Shanghai 201315, China

*jtsheng@lps.ecnu.edu.cn; hbwu@phy.ecnu.edu.cn

**Clock synchronization is critically important in positioning, navigation and timing systems. While its performance has been intensively studied in a wide range of disciplines, much less is known for the fundamental thermodynamics of clock synchronization–what limits the precision and how to optimize the energy cost for clock synchronization. Here, we report the first experimental investigation of two stochastic clocks synchronization, unveiling the thermodynamic relation between the entropy cost and clock synchronization in an open cavity optomechanical system. Two autonomous clocks are synchronized spontaneously by engineering the controllable photon-mediated dissipative optomechanical coupling and the disparate decay rates of hybrid modes. The measured dependence of the degree of synchronization on entropy cost exhibits an unexpected non-monotonic characteristic, indicating that the perfect clock synchronization does not cost the maximum entropy and there exists an optimum. The investigation of transient dynamics of clock synchronization exposes a trade-off between energy and time consumption. Our results reveal the fundamental relation between clock synchronization and thermodynamics, and have a great potential for precision measurements, distributed quantum networks, and biological science.**

Clock (time) synchronization [1] is essential to numerous scenarios, ranging from distributed systems to biological systems, including precision navigation [2] as well as quantum key distribution [3]. In spite of significant achievements made in clock synchronization across various fields, for example, the synchronization of distant optical clocks at the femtosecond level [2], less is known about the fundamental aspect of clock synchronization, i.e. the precision limit and energy constraints. To address these questions, it has to find its connection to thermodynamics, which remains elusive so far. Recently, it has been predicted that the synchronization of molecular clocks requires extra energy cost [4], whilst the study of thermodynamic cost of synchronizing cilia arrays argued that the total dissipation is reduced upon synchronization [5].

Optomechanics, which has made tremendous progress in exploring quantum physics at macroscopic scales and ultrasensitive metrology [11], is recognized as an excellent platform for studying nonequilibrium thermodynamics [12-20]. Synchronization and entrainment have been demonstrated in such systems [21-25]. Recently, the thermodynamic costs of timekeeping has explored by investigating a nanoscale clock in an optomechanical system [26]. Despite numerous theoretical studies have been devoted to the thermodynamics aspects of synchronization with various models [6-10], to our knowledge, no experimental demonstration has been performed to investigate the relationship between the clock synchronization and thermodynamics, which is mainly due to the challenges in developing an ideal system that can perform clock synchronization and enable the precision measurement of thermodynamic costs.

In this work, we report an experimental study of the thermodynamic cost of clock synchronization. Clocks are nontrivial devices of measuring time, which are open nonequilibrium irreversible systems [27-31]. Here, two spatially separated nanomechanical membranes are used as autonomous clocks which are stochastically driven independently. These two membrane resonators are placed inside an optical cavity, which consist of a membrane-in-the-middle optomechanical system [32,33]. Thereby, the interaction between nanoscale clocks can be flexibly controlled through a common cavity field via dynamical backaction [11]. The clocks are spontaneously synchronized by increasing the effective coupling strength over a critical value (see Fig. 1a). The degree of synchronization is enhanced as the coupling strength increases until saturated at the maximum value. The mechanism of the synchronization state is from the disparate decay rates of hybrid modes cause by the dissipative coupling [34-36], which is also known as the transient synchronization [37]. This is in contrast to the previous demonstrations in optomechanics [22-25], which are based on phonon lasers, i.e., the limited cycle in nonlinear systems. By monitoring the single trajectories in real-time, the entropy production is assessed by considering the three-body interacting scheme in the nonequilibrium steady state (NESS), which fully takes optical mode and phononic modes simultaneously including the thermal baths into account. We find that a minimum thermodynamic cost is required to achieve the clock synchronization. Importantly, the degree of synchronization as a function of entropy cost shows a surprisingly non-monotonic characteristic, implying that the perfect clock synchronization does not always cost the maximum entropy. In addition, the transient dynamics of synchronization is investigated with respect to the transient correlation and the transient entropy flux, which shows that more dissipation is required in order to synchronize the clocks more quickly.

Our experiment is performed in a two-membrane-in-the-middle optomechanical system [17]. Two spatially separated silicon nitride nanomechanical membranes are placed inside a Fabry-Perot cavity. The fundamental vibrational modes of membranes are used in the experiment, which are nearly degenerate with natural frequencies $\omega_{1,2} \sim 2\pi \times 400 \text{kHz}$ and different mechanical decay rates $\gamma_{1(2)} \approx 2\pi \times 7(14) \text{Hz}$. The motions of membranes are monitored by two weak probe laser fields separately [25]. The cavity field interacts with two membrane resonators simultaneously via radiation pressure, which provides an effective coupling between two membranes.

Two membranes are driven independently by the white noises as two autonomous clocks, which operate like any thermodynamic machines by consuming a resource and creating waste in the form of entropy, and simultaneously output a train of ticks [26]. Without the intracavity field, two clocks are unsynchronized. Figure 1b displays the oscillatory real-time trajectories of displacements $x_{1,2}$ of two membrane clocks, allowing the identification of the ticks of thermodynamic clocks. The corresponding phase portrait exhibits a random and isotropic structure, which indicates that two clocks are not synchronized (Fig. 1c). The unsynchronized state can also be identified as the distinct power spectra of two clocks with an intrinsic frequency mismatching $\Delta\omega = \omega_1 - \omega_2 \sim 2\pi \times 200 \text{Hz}$ (Fig. 1d). When the intracavity field is increased over a synchronization threshold, two clocks become completely synchronized (Fig. 1e). The synchronized clocks with near equal oscillated frequency are anisotropic in the phase portrait and show the strong correlation (Figs. 1f-1g).

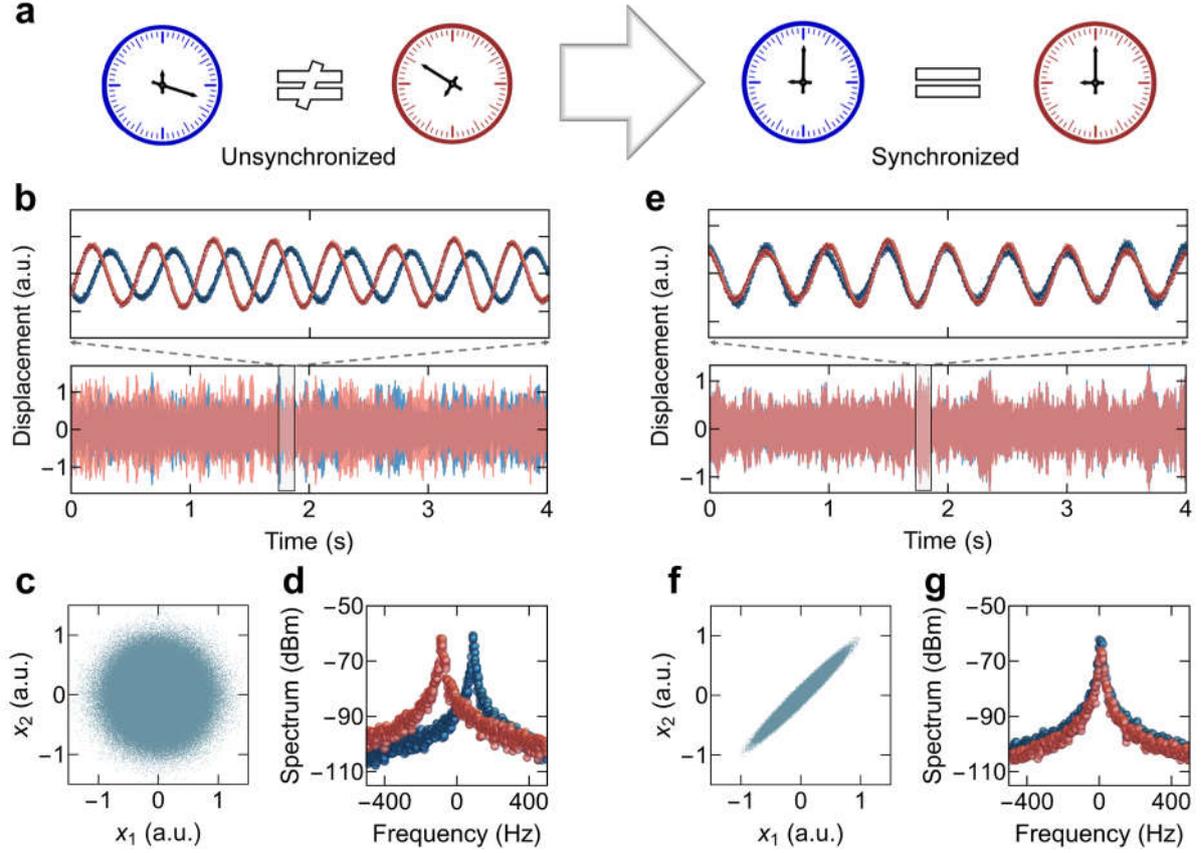

**Fig. 1. Synchronization of clocks via interacting with optomechanically induced dissipative coupling.** (**a**) Illumination of the transition of two unsynchronized thermodynamic clocks to the synchronized state by increasing the effective dissipative coupling. (**b-d**) Measured real-time trajectories of displacements $x_{1,2}$, phase portrait ($x_1$ vs $x_2$), and power spectra of two autonomous clocks in the non-synchronized state, respectively. (**e-g**) Measured real-time trajectories of displacements, phase portrait, and power spectra of two autonomous clocks in the synchronized state, respectively. The zoomed-in views in Figs. 1b and 1e display the trajectories in a duration of 20 μs. The blue and red curves in Figs. 1b, 1d, 1e, and 1g represent two clocks, respectively.

The interaction Hamiltonian of such a composite system of two membranes interacting with a common cavity field is $\hat{H}_{\text{int}} = -\sum_{i=1,2} g_i \hat{a}^\dagger \hat{a} (\hat{b}_i^\dagger + \hat{b}_i)$, where $\hat{a}$ and $\hat{b}_i$ are the annihilation operators for optical and phononic modes, respectively. $g_i$ is the single-photon optomechanical coupling rate between the optical and phononic modes. For better understanding the mechanism of clock synchronization, the system can be approximated to a coupled-mode model with an effective interacting Hamiltonian $\hat{H}_{\text{int}}^{\text{eff}} = \Lambda(\hat{b}_1^\dagger \hat{b}_2 + \hat{b}_2^\dagger \hat{b}_1)$ after eliminating the cavity field [17]. Here $\Lambda = G_1 G_2 \chi_c$ is the effective coupling rate between two membranes, $G_{1,2} = 2g_{1,2}\sqrt{\bar{n}_{cav}}$ is the effective optomechanical interaction enhanced by the intracavity mean photon number $\bar{n}_{cav}$. Two membranes are engineered to have the same effective optomechanical coupling strength with the opposite sign, i.e., $G = G_1 = -G_2$. $\chi_c$ is the effective intracavity field susceptibility (see Supplementary Note 1). Thus, the effective coupling rate can be rewritten as $\Lambda = \delta + i\Gamma$ with the coherent coupling $\delta = -G^2 \text{Re}(\chi_c)$ and the dissipative coupling $\Gamma = -G^2 \text{Im}(\chi_c)$ between two phononic modes.

The eigenvalues of such a radiation-pressure mediated coupled phononic system are obtained as $\lambda_\pm = (\Delta\omega/2 - i(\gamma_1 + \gamma_2 + 4\Gamma)/4) \pm \sqrt{(\Delta\omega - i(\gamma_1 - \gamma_2)/2)^2/4 + (\delta + i\Gamma)^2}$, where the real and imaginary parts correspond to the eigenfrequencies ($\omega_\pm = \text{Re}(\lambda_\pm)$) and linewidths ($\gamma_\pm = -2\text{Im}(\lambda_\pm)$) of the normal modes, respectively. Consequently, the normal mode $\lambda_+$ with the smaller decay rate $\gamma_+$ is dominant in the long-time dynamics after a transient time. Although both dissipative and coherent coupling terms are considered, it is important to emphasize that the synchronization relies on the dissipative coupling, which leads to the disparate decay rates of hybrid modes [36]. In contrast, the coherent coupling only induces the mode splitting [17]. The clock synchronization mechanism discovered here is similar to the transient synchronization proposed for quantum synchronization, where the coherent coupling and common bath are used instead [37]. Another prominent feature of such a mechanism of synchronization is that it can occur in an effective linear system, in contrast to the fact that synchronization is typically recognized as a universal phenomenon in nonlinear science [1].

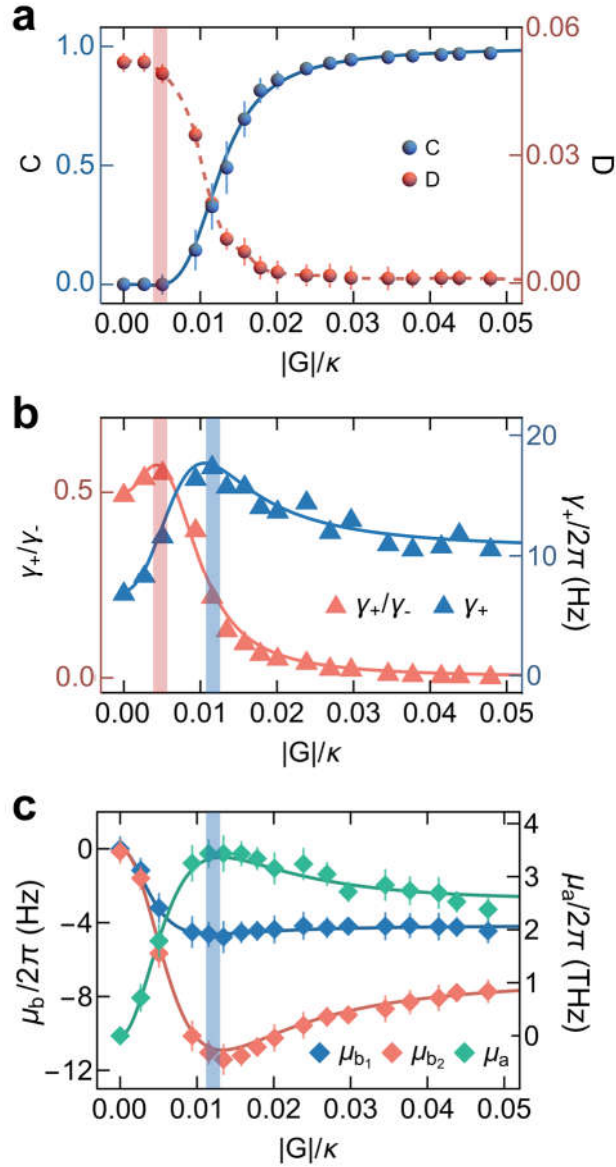

**Fig. 2. Quantitative analysis of clock synchronization and the entropy production rate.** (**a**) Measured degree of synchronization $C$ (blue dots) and deviation of clock synchronization $D$ (red dots) as a function of $|G|/\kappa$. The blue solid and red dashed curves are the corresponding theoretical simulations. (**b**) The ratio of normal mode linewidths $\gamma_+/\gamma_-$ (red triangles) and the normal mode linewidth $\gamma_+$ (blue triangles) as a function of $|G|/\kappa$. (**c**) Measured entropy production rates contributed by the phononic ($\mu_{b_{1,2}}$) and photonic ($\mu_a$) modes as a function of $|G|/\kappa$. The blue and red diamonds are for the phononic modes and the green diamonds are for the photonic mode, respectively. The error bars are the standard deviations. The blue, red and green curves are the corresponding theoretical simulations.

The experimental measurement of the degree of synchronization as a function of $|G|/\kappa$ ($\kappa \approx 2\pi \times 2\text{MHz}$ is the cavity decay rate) is presented in Fig. 2a. The degree of synchronization is defined as the Pearson correlation coefficient of time-dependent displacements in the NESS, i.e. $C \equiv \langle \delta x_1 \delta x_2 \rangle / \sqrt{\langle \delta x_1^2 \rangle \langle \delta x_2^2 \rangle}$ [38]. Here $\delta x_{1,2} = x_{1,2} - \langle x_{1,2} \rangle$ is the fluctuation of displacement, and $\langle \rangle$ denotes the average over a time interval. Without the intracavity field ($|G|= 0$), two clocks oscillate independently driven by the uncorrelated thermal baths, and thus $C = 0$. As $|G|/\kappa$ increases, C keeps around zero until above a critical coupling strength (depending on $\Delta\omega$ and $\chi_c$), i.e., $|G_c|/\kappa \sim 0.005$, indicated by the red translucent line in Fig. 2a. Beyond $|G_c|/\kappa$, C increases quickly as $|G|/\kappa$ is further enhanced and eventually is saturated to the maximum value of synchronization ~ 1, where two clocks are very close to the state of perfect in-phase synchronization.

In order to quantitatively understand the behavior of the degree of synchronization as a function of |G|/κ, the ratio of γ₊/γ₋ is plotted in Fig. 2b, where the red triangles are the experimental data and the red curve is the theory. As one can see in Fig. 2b, γ₊/γ₋ remains essentially unchanged before the critical coupling strength $|G_c|$, while it decays rapidly after $|G_c|/\kappa$. As a result, two clocks are synchronized to the normal mode with the smaller decay rate $\gamma_+$ as γ₊/γ₋ approaches to zero, at the expense of the faster decay of the other normal mode. The divergence of γ₊ and γ₋ is attributed to the dissipative coupling, since the dissipative coupling causes the damping repulsion and level attraction [36], while the coherent coupling leads to the contrary phenomenon, i.e. level avoided crossing. It is worth mentioning that the opposite sign of optomechanical coupling strength, i.e. $G_1 = -G_2$, leads to the in-phase synchronization, while the same optomechanical coupling strength ($G_1 = G_2$) leads to the out-of-phase synchronization.

We also characterize the timekeeping error of clock synchronization, as shown in Fig. 2a. Compared to the accuracy quantified the timekeeping error for a single clock defined as $N_{1,2} = t_{1,2}^2 / \Delta t_{1,2}^2$, where $t_{1,2}$ is the period of clock 1 and 2 (the interval between successive ticks) and $\Delta t_{1,2}$ is the standard deviation of the interval [26,30], here the deviation D of clock synchronization is defined as $D \equiv \langle (\tau - \langle \tau \rangle)^2 \rangle / \langle (t_1 + t_2)/2 \rangle^2$, where $\tau = t_2 - t_1$ is the period difference of two clocks. Typically, $\langle (t_1+t_2)/2 \rangle^2$ is approximately a constant and thus D is mainly determined by the fluctuation of $\tau$. Under the critical coupling strength $|G_c|/\kappa$, D is relatively large, indicating that τ fluctuates with a relatively large value, which depends on the intrinsic frequency mismatching and linewidths of two clocks. While, beyond $|G_c|/\kappa$, D decreases until close to zero, where the long-lived normal mode dominates in two clocks in the synchronized state. This means that the timekeeping of the two clocks is coordinated almost exactly the same way as the coupling strength is strong enough.

To determine the relation between the clock synchronization and thermodynamic cost, the entropy production rate of the entire system has to be assessed. The optical mode cannot be ignored, since it contributes predominately to the entropy production rate of clock synchronization compared to the phononic modes. According to the second law of thermodynamics, the rate equation for the variation of the entropy S can be formulated as $dS/dt = \Pi(t) - \Phi(t)$, where $\Pi(t)$ is the irreversible entropy production rate and $\Phi(t)$ is the entropy flux rate from the system to the environment [39]. By utilizing the Shannon entropy of the Wigner function to calculate the entropy S of the system [14], the total irreversible entropy production rate of the system in the NESS is obtained as (see Supplementary Note 2)

$$\Pi_s = \Phi_s = \gamma_1 \left( \frac{\bar{n}_{b_1}^{eff}+1/2}{\bar{n}_{b_1}^{th}+1/2} - 1 \right) + \gamma_2 \left( \frac{\bar{n}_{b_2}^{eff}+1/2}{\bar{n}_{b_2}^{th}+1/2} - 1 \right) + 2\kappa \bar{n}_a^{eff} = \mu_{b_1} + \mu_{b_2} + \mu_a. \quad (1)$$

Here $\Pi_s$ and $\Phi_s$ are denoted as the quantities in the NESS, and the entropy production rate $\Pi_s$ generated in system equals the entropy flux rate $\Phi_s$ to overall thermal reservoirs in the NESS. $\mu_{b_{1,2}}$ and $\mu_a$ represent the contributions of the phononic and photonic modes to $\Pi_s$, respectively. $\bar{n}_{b_{1,2}}^{th} \approx k_B T_{1,2}/\hbar\omega_{1,2}$ is the average phonon occupation in the thermal equilibrium state interacting only with individual thermal bath. $\bar{n}_a^{eff}$ and $\bar{n}_{b_{1,2}}^{eff}$ are the effective mean photon and phonon numbers in the NESS. Equation (1) manifests the entropy cost attributing to the fluctuations that the system must pay to sustain its NESS and directly quantifies the irreversibility of the driven-dissipative dynamics of three coupled resonators.

In the experiment, $\mu_{b_{1,2}}$ is directly obtained by measuring the effective mean phonon number $\bar{n}_{b_{1,2}}^{eff} = \langle \hat{b}_{1,2} \hat{b}_{1,2}^\dagger \rangle$. To determine $\mu_a$, the cross-correlation between two phononic modes $\bar{n}_{cross}^{eff} = \langle (\hat{b}_1 + \hat{b}_1^\dagger)(\hat{b}_2 + \hat{b}_2^\dagger) \rangle / 2$ is required (see Supplementary Note 2). Figure 2c displays the experimental measurements of $\mu_{b_{1,2}}$ and $\mu_a$ as a function of $|G|/\kappa$. Both $\mu_{b_1}$ and $\mu_{b_2}$ have negative values, implying that the flux of the entropy flows from the phononic modes to the photonic mode, as a signature of the optomechanical cooling effect [14]. As the coupling strength increases, $|\mu_{b_{1,2}}|$ first increases, corresponding to the lowering of the effective temperature of the phononic modes owing to the enhanced cooling strength, and then reduces at a turning point $|G_t|/\kappa \sim 0.013$, denoted by the blue translucent line in Figs. 2b and 2c, which is resulted from the dark-mode effect [40] that suppresses the optomechanical cooling of the long-lived normal mode. The dark mode effect is clearly observed in Fig. 2b, where the value of $\gamma_+$ ceases to increase beyond the turning point and reduces as $|G|/\kappa$ further enhances.

The negative entropy of phononic modes results from the cost of the positive entropy of the optical mode. Therefore, $\mu_a$ has the opposite behavior compared to $\mu_{b_{1,2}}$, i.e. $\mu_a$ increases at a relatively small value of $|G|/\kappa$ and then decreases after the turning point, which is illustrated by the green diamonds in Fig. 2c. According to the measurements shown in Fig. 2c, the value of $\mu_a$ is on the order of terahertz (THz), which is much larger than the values of $\mu_{b_{1,2}}$. Consequently, the total entropy production rate of the system $\Pi_s$ is mostly contributed from the photonic mode and has a positive value in the NESS, which satisfies the second law of thermodynamics.

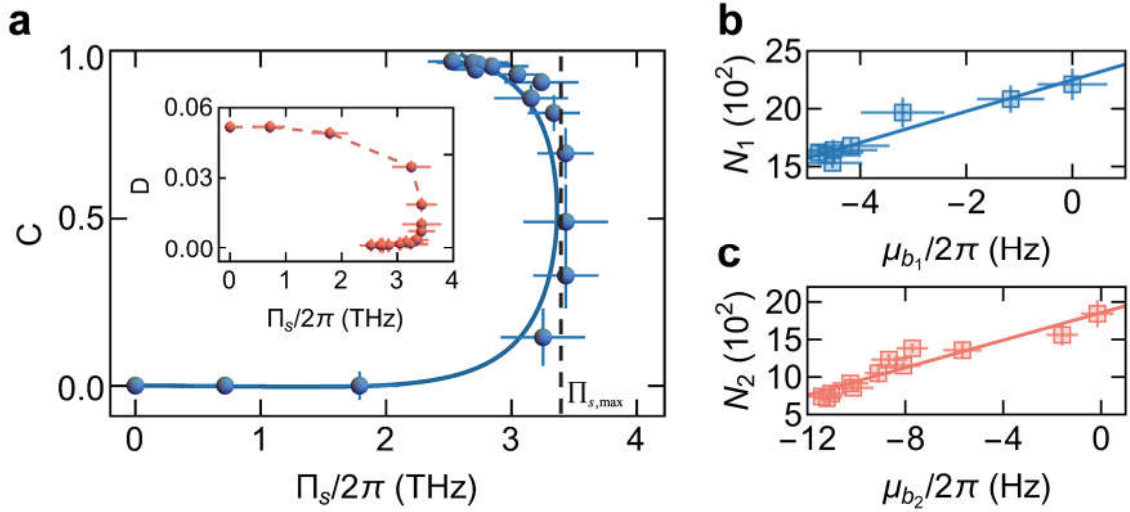

**Fig. 3. Thermodynamic relation.** (**a**) Degree of synchronization (C) as a function of total entropy production rate $\Pi_s$ in the NESS. The inset shows the deviation of clock synchronization (D) as a function of $\Pi_s$. The blue and red dots are the experimental measurements and the blue solid and red dashed lines are the corresponding theoretical simulations. The error bars are the standard deviations. (**b,c**) Accuracy of single clock versus the entropy production rate of the corresponding the phononic mode.

We investigate the relationship between thermodynamic cost and clock synchronization, as shown in Fig. 3. The degree of synchronization C is presented as a function of total entropy production rate $\Pi_s$ in the NESS. As $\Pi_s$ increases, C is nearly zero until beyond the critical point. Surprisingly, C is not a monotonically increasing function of $\Pi_s$. The turning point in Fig. 3 is resulted from the dark mode effect. Consequently, the optimized degree of synchronization can be achieved at a smaller value than the maximum entropy production rate $\Pi_{s,\max}$ (indicated by the black dashed line in Fig. 3). The non-monotonic characteristic between the clock synchronization and the entropy production rate is a new finding which can maximize clock synchronization under certain specific thermodynamic cost. The inset of Fig. 3 shows the deviation of synchronization $D$ as a function of $\Pi_s$, which also shows a non-monotonic behavior. Beyond the turning point, the deviation of synchronization can approach nearly the minimum of timekeeping error of two clocks ~ 0, i.e., approximately identical clock coordination with a relative smaller thermodynamic cost. On the contrary, the accuracy $N_{1,2}$ of single clock exhibits a linear relation as a function of the corresponding entropy production rate $\mu_{b_{1,2}}$, as shown in Figs. 3b and 3c, respectively. Such a linear relationship has been recognized as being universal in both quantum and classical regimes [26,30].

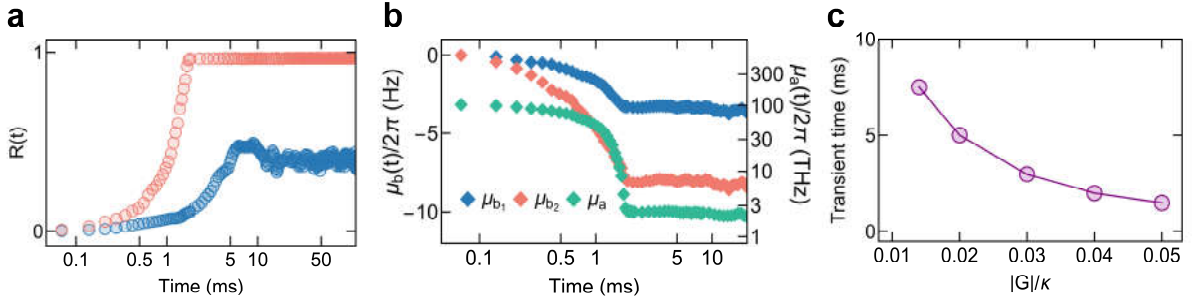

**Fig. 4. Transient dynamics of clock synchronization.** (**a**) Transient correlation (R) as a function of time at different coupling strengths. The blue and red circles are for |G|/κ ~ 0.01 and 0.04, respectively. (**b**) Transient entropy flux as a function of time for |G|/κ ~ 0.04. Each curve is averaged over 600 transient time trajectories. (**c**) Transient time as a function of |G|/κ.

Last, the transient dynamics of clock synchronization is investigated. The transient correlation is defined as $R(t) \equiv \sum_{j=1}^{N}\left(\delta x_{1,j}(t)\delta x_{2,j}(t)\right)/\sqrt{\sum_{j=1}^{N}\delta x_{1,j}^{2}(t)\sum_{j=1}^{N}\delta x_{2,j}^{2}(t)}$ ($\sum_{j=1}^{N}$ represents the sum of N transient trajectories). The transient correlations at different coupling strengths are presented in Fig. 4a, respectively. The system is quenched at zero time by switching on the cavity field. R(t) evolves from zero to a high value (~ 1) for a relatively large $|G|/\kappa$ ~ 0.04, while R(t) exhibits an "overshoot" behavior for $|G|/\kappa$ ~ 0.01 [41]. The transient entropy fluxes $\mu_{b_{1,2}}(t)$ and $\mu_{a}(t)$ decrease with time and then remain to a stationary state, as shown in Fig. 4b. The transient time is defined as the time interval before R(t) reaches the maximum value. As one can see in Fig. 4c, the larger $|G|/\kappa$ is, the smaller the transient time is. This is due to the fact that the larger $|G|/\kappa$ leads to the larger difference between γ₊ and γ₋, which causes the system to reach the maximum degree of synchronization more quickly.

In conclusion, we have realized the synchronization of two thermomechanical clocks via the radiation pressure induced coupling in a multimode optomechanical system. The clock synchronization is achieved due to the disparate decay rates of hybrid modes resulting from the cavity-induced dissipative interaction. More importantly, the thermodynamic relation between the clock synchronization and the entropy production rate is experimentally extracted. The degree of synchronization as a function of the entropy production rate in the NESS presents an unexpected non-monotonic characteristic. Our experimental results reveal a fundamental relation between clock synchronization and thermodynamics, which can lead to a deep understanding of thermodynamics of clock synchronization and open up significant opportunities for multiple disciplines.